\documentclass{article}

\usepackage{arxiv}

\usepackage[utf8]{inputenc} % allow utf-8 input
\usepackage[T1]{fontenc}    % use 8-bit T1 fonts
\usepackage{hyperref}       % hyperlinks
\usepackage{url}            % simple URL typesetting
\usepackage{booktabs}       % professional-quality tables
\usepackage{amsfonts}       % blackboard math symbols
\usepackage{nicefrac}       % compact symbols for 1/2, etc.
\usepackage{microtype}      % microtypography
\usepackage{cleveref}       % smart cross-referencing
\usepackage{lipsum}         % Can be removed after putting your text content
\usepackage{graphicx}
\usepackage{doi}
\usepackage{fancyvrb}
\usepackage{enumitem}

\usepackage{algorithm}% http://ctan.org/pkg/algorithms
\usepackage{algpseudocode}% http://ctan.org/pkg/algorithmicx

\usepackage{multicol}

\usepackage{tabularx}
\usepackage{capt-of}

\title{A Comparative Analysis of Port Scanning Tool Efficacy}

\author{
    Jason M. Pittman \\
    ORCID: 0000-0002-5198-8157 \\
}

\hypersetup{
    pdftitle={A Comparative Analysis of Port Scanning Tool Efficacy}
    pdfsubject={}
    pdfauthor={Jason M. Pittman}
    pdfkeywords={port scanning, cybersecurity, experiment, tools and techniques, efficacy}
}

\begin{document}
\maketitle

\begin{abstract} 

Port scanning refers to the systematic exploration of networked computing systems. The goal of port scanning is to identify active services and associated information. Although this technique is often employed by malicious actors to locate vulnerable systems within a network, port scanning is also a legitimate method employed by IT professionals to troubleshoot network issues and maintain system security. In the case of the latter, cybersecurity practitioners use port scanning catalog exposed systems, identify potential misconfigurations, or test controls that may be running on a system. Existing literature has thoroughly established a taxonomy for port scanning. The taxonomy maps the types of scans as well as techniques. In fact, there are several tools mentioned repeatedly in the literature. Those are Nmap, Zmap, and masscan. Further, the presence of multiple tools signals that how a port scanner interacts with target systems impacts the output of the tool. In other words, the various tools may not behave identically or produce identical output. Yet, no work has been done to quantify the efficacy for these popular tools in a uniform, rigorous manner. Accordingly, we used a comparative experimental protocol to measure the accuracy, false positive, false negative, and efficiency of Nmap, Zmap, and masscan. The results show no difference between port scanners in general performance. However, the results revealed a statistically significant difference in efficiency. This information can be used to guide the selection of port scanning tools based on specific needs and requirements. As well, for researchers, the outcomes may also suggest areas for future work in the development novel port scanning tools.

\end{abstract}


%% start two column
\begin{multicols*}{2}

\section{Introduction} 
Many cybersecurity tools and techniques have two use cases. One on hand, these tools and techniques can be used to validate system or network configurations, map services and architectures, as well as aid troubleshooting. On the other hand, the same can be used by malicious actors to discover endpoints and extract valuable information to feed into exploitation chains. Port scanners firmly reside in this set of two-sided use cases \cite{de1999review, el2011evaluation, Yuan_2020}. 

Port scanners, as a tool, leverage RFC \cite{comer1994probing} standards for TCP/IP to interact with computing endpoints. It is important to understand two points here. First, such interactions are governed by the RFC standards. For example, RFC 1180 \cite{rfc1180} is the standard for TCP/IP and serves as an underlying structure for associated internetworking protocols. One such protocol, used extensively by port scanners, is TCP \cite{rfc793}. The techniques of port scanning layer various elements of these standards to enumerate the endpoints and the services running on them \cite{de1999review}. More specifically, port scanners can identify open ports, \cite{Staniford2002Scan}, fingerprint service and operating system information \cite{lyon1998}, and assess firewall configurations \cite{Zhang_2015}. A majority of techniques manipulate TCP flags (e.g., SYN, ACK, FIN) to elicit the endpoint responses. Some techniques exist, though, which use UDP or IP.

While a healthy foundation of port scanning literature exists, there has been little examination of port scanner efficacy \cite{im2016performance, Yuan_2020}. To that end, we found two out of 12 studies between 1994 and 2022 demonstrating a quantitative evaluation. Moreover, no study comparatively analyzed port scanners for accuracy, false positives, false negatives, and efficiency (i.e., \textit{efficacy}). This limits research innovation and leaves practitioners without clear evidence of which tool may be best suited for certain operational conditions.

For that reason, the purpose of this study was to quantitatively compare the efficacy of three port scanners. We used an experimental design to generate scan metrics, collect system utilization data during the scans, and employ statistical analysis. This work has practical significance for cybersecurity practitioners. Since port scanning is a legitimate practice for network defense, selecting an accurate tool with few or no false positives or false negatives may be important to network and system operators. At the same time, a variety of significant use cases exist for researchers. Understanding existing port scanners may lead to development of new port scan algorithms and prototypes. Further, the same understanding may inform development of robust port scan detection mechanisms.

The rest of this work is organized in four sections. First, we present related work on port scanning as a means of establishing a conceptual framework. Afterward, we detail the research method including the procedure we followed. Then, we show the results of the data analysis and offer comparative insights. We end by detailing a series of recommendations and ideas for future work based on results informed conclusions.

\section{Related Work} 
The literature background for port scanning is not extensive. However, existing research has established a robust conceptual framework to situate our work in. Specifically, because port scanning techniques are inherently \textit{technical}, we want to calibrate important terminology and technical details. Moreover, understanding what efficacy data exists is important foundational material. 

\subsection{Port Scanning Foundation} 

The term \textit{port scanning} has its roots in the early days of computer networking and can be traced back to the late 1980s and early 1990s. As the Internet expanded and became more prevalent, network administrators and security professionals required tools to gain insight into the state of their networks. Identifying active network services, as well as open and closed ports on hosts, was a crucial task for these professionals. This process came to be known as port scanning and has since become a vital technique for maintaining network security.

Port scanning leverages TCP/IP capabilities to identify computing systems within a network. As network protocols utilize distinct ports, a thorough scan of a broad range of ports is crucial for comprehensive information gathering. The maximum number of ports that can be scanned is 65535, classified into three categories: well-known ports (0-1023), registered ports (1024-49151), and dynamic or private ports (49152-65535) \cite{rfc814, rfc6335}.

For illustrative purposes, let us consider TCP-based port scanning. Being connection-oriented, TCP communications rely on an initial four-way handshake (Algorithm 1). In the context of port scanning, the Server response in step 2 signals (a) the endpoint is online and (b) an active service is available on the port. Additionally, attention should be given to the TCP flags (SYN and ACK) present in the exchange. Client and Server use these flags to control the networking exchange \cite{comer1994probing}.

\begin{algorithm}[H]
\textbf{Client} $\rightarrow$ \textbf{Server}: SYN (Seq=x) 

\textbf{Server} $\rightarrow$ \textbf{Client}: SYN (Seq=y, Ack=x+1) 

\textbf{Client} $\rightarrow$ \textbf{Server}: ACK (Seq=a+1, Ack=b+1) 

\textbf{Server} $\rightarrow$ \textbf{Client}: ACK (Seq=b+1, Ack=a+1) 
\caption{TCP Four-Way Handshake}
\end{algorithm}

TCP is but one means of transport and thus one means of port scanning. While one scan technique exists that uses the full handshake \cite{lyon1998, de1999review}, the majority either manipulate the handshake sequence (i.e., sending an early connection reset), manipulate TCP flags, or leverage other packet fields in unexpected ways. Although the TCP handshake is based on SYN and ACK, there are four additional flags available: URG, PSH, RST, and FIN \cite{rfc814, rfc6335}. Additionally, there are \textit{fragment offset} and \textit{options} open to modifications. The standard define all of these in the IPv4 packet structure (Figure 1).

% IPv4 packet
\begin{figure}[H]
	\setlength\abovecaptionskip{-0.7\baselineskip}
	\centering
	\includegraphics[width=\linewidth]{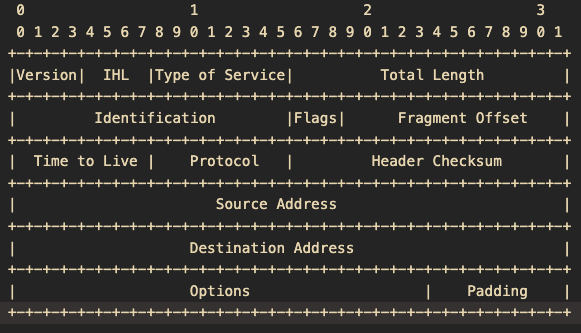}
	\caption{The fields within an IPv4 packet and their relative sizes in bits}
	\label{fig:ipv4}
\end{figure}

\subsection{Port Scanning Conceptual Framework}
Port scanning was made well-known in the literature by Fyodor \cite{lyon1998}. This work introduced the concept of automated network mapping including a method for identifying the operating system of a remote host. The tool, Nmap, sends TCP/IP probes to specific ports and analyzes the responses. The paper explained how the behavior of the TCP/IP stack, as well as responses to different types of probes, such as initial sequence numbers (ISNs) and options in TCP headers, could be used to determine service states as well as the operating system. The work also described the use of this technique to fingerprint endpoints as well as the challenges and limitations with port scanning. Nmap features heavily throughout port scanning literature, popular cybersecurity culture, and a variety of media outputs. 

Now, Fyodor \cite{lyon1998} was at least conceptually based on earlier work by Comer and Lin \cite{comer1994probing}. The authors conducted a series of experiments to measure the performance of TCP implementations. The experiments involved sending a variety of TCP probes to the target systems and observing the responses. Such \textit{active probing} is port scanning in function if not also in form. Furthermore, a critical takeaway is the idea that active probing makes no assumptions about target endpoints. It is the responses from the endpoints which reveal all the information. 

De Vivo et al. \cite{de1999review} generalizes from the port scanning foundation provided in Fyodor \cite{lyon1998} and several \cite{comer1994probing, wright1995tcp} others. The significance of De Vivo et al. \cite{de1999review} emerges from the rigorous classification applied to port scanning techniques and procedures. The paper described the different types of port scans, such as TCP connect scans and SYN scans as \textit{classical}. This is in relation to \textit{indirect} and \textit{stealth} scanning. The latter is also referred to as a FIN, XMAS, or NULL scan. The former is realized by bouncing scans off of a zombie endpoint. The work goes on to describe scanning techniques. These includes \textit{decoy} scanning, \textit{fragmented} scanning, and \textit{coordinated} or \textit{distributed} scanning, UDP scanning, and ICMP sweeping. 

Staniford et al. \cite{Staniford2002Scan} detailed several port scanning concepts in the process of developing scan detection mechanisms. \textit{Footprint} encompasses the networking concept of a socket. That is, an IP address paired with a TCP or UDP port. The authors then make a distinction between that and a port scanning \textit{script}. Here, a script is the time associated with probing a footprint. Together, footprint and script converge to an active probing technique Staniford et al. refer to as \textit{horizontal scanning}. Horizontal scanning, or probing for a single service across a network segment, is a concept picked up in later work \cite{barnett2008towards, laroche2009evolving, Bhuyan2011Survey}. 

Barnett et al. \cite{barnett2008towards} introduced a system for categorizing network scanning techniques, which is significant in establishing a structured and comprehensive classification of these methods and their applications. The authors proposed a taxonomy that classifies network scanning techniques according to the level of interaction with the target system, the information gathered, and the purpose of the scan. This taxonomy builds on the work of De Vivo et al. \cite{de1999review}, incorporating additional types and techniques. To do so, the authors employed a multifaceted approach that involved generating scan traffic in a lab setting using Nmap. Additionally, Barnett et al. established a network telescope to capture traffic in the wild. 

The authors identified a taxonomy of network scanning techniques consisting of seven different scans categorized into three levels. The root node of the taxonomy is the TCP/IP scan, which is followed by layer 2, layer 3, scanning speed, and scan distribution. Notably, scan distribution extends \cite{Staniford2002Scan} by adding the concept of \textit{vertical} scans to horizontal. The layer 3 category is further divided into ICMP, TCP, and UDP. The TCP category includes scans using SYN, ACK, and FIN flags. The scanning speed is classified into three categories, namely, slow, medium, and rapid, while the scan distribution is defined by the relationship between the source and destination, such as one-to-one, one-to-many, many-to-one, and many-to-many. 

Bhuyan et al. \cite{Bhuyan2011Survey} described general approaches to port scanning by extended prior work by De Vivo et al. \cite{de1999review}, Staniford et al. \cite{Staniford2002Scan}, and Fyodor \cite{lyon1998}. The authors \cite{Bhuyan2011Survey} made an important differentiation between single-source and distributed port scans. Organizationally, single-source scanning encompass \textit{types} (vertical, horizontal, strobe, and block), \textit{port} (some or all, single, multiple, and all), and \textit{target} (single, multiple IPs, multiple IPs, and multiple IPs). Distributed scans are many to one or many to many. 

An additional contribution can be found in how the Bhuyan et al. detail port scanning techniques, including the source and target TCP/IP interactions. In doing so, the authors demonstrate how the TCP/IP standards are utilized in port scanning. The implication being that port scan enumeration is unavoidable given the fundamental rules defined in such standards. A synthesis of this contribution is in Figure 2.

\begin{figure}[H]
	\setlength\abovecaptionskip{-0.7\baselineskip}
	\centering
	\includegraphics[width=\linewidth]{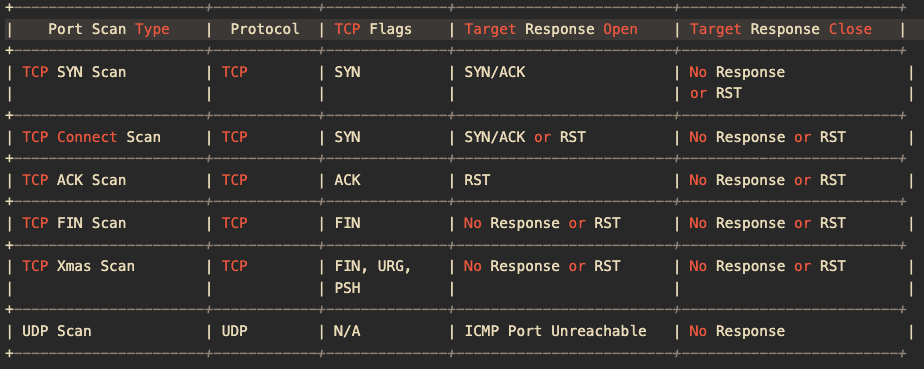}
	\caption{The various port scan techniques and associated protocol interactions.}
	\label{fig:scans}
\end{figure}

Similarly, El-Nazeer and Daimi \cite{el2011evaluation} evaluated network port scanning tools with the goal of identifying the best tool for network administrators to use to protect their networks. The authors used a comparative analysis research method to evaluate eight port scanning tools. Nmap and unicornscan were in the port scanner evaluation set and are featured repeatedly in later studies. El-Nazeer and Daimi described 15 evaluation criteria ranging from the ability to perform SYN scan to whether the tool is free and open source or not. They lab tested the port scanners for 13 of the criteria and used documentation to derive values for the other two. Somewhat puzzling given the lab testing is the lack of quantitative measures. In place of such data, the authors qualitatively declare Nmap as having the most robust features amongst the examined port scanners. 

Bou et al. \cite{bou2013cyber} provided a detailed examination of the various types of cyber scanning techniques employed to identify different features of networks. The authors categorized port scanning techniques into two main categories: passive and active. Passive scanning techniques involve listening to network traffic to collect information about the target network without sending any packets. On the other hand, active scanning techniques entail sending packets to a target host to elicit a response, which can help determine the host's characteristics and identify vulnerabilities. The latter harkens back to Comer and Lin \cite{comer1994probing} explicitly and to others such as Staniford et al. \cite{Staniford2002Scan} implicitly.

Furthermore, Bou et al. \cite{bou2013cyber} added to the organizational structure provided by De Vivo et al. \cite{de1999review} and Barnett et al. \cite{barnett2008towards} but differed in semantics. For instance, the De Vivo et al. classical, indirect, and stealth scans map under the nature of active and passive scanning offered by Bou et al. \cite{bou2013cyber}. As well, the semantic developed by \cite{barnett2008towards} around relations between scanner and target (e.g., one-to-many) falls under \textit{approarch} in Bou et al. \cite{bou2013cyber}.  Bou et al. also offered \textit{strategy} as a way to categorize directional relationship between scanner and target. This continued with Bou et al. defining \textit{approach} as aim and method. The former having a wide range of targets and latter having specific targets. Method encompasses  single source or distributed. Then, \textit{techniques} includes open, half open, stealth (SYN, ACK, IDLE, FIN, XMAS, NULL, ACK Window, and Fragmentation), sweep (ICMP, SYN) and miscellaneous (FTP Bounce, UDP, IP, and RPC).

Kumar and Sudarson \cite{kumar2014innovative} and Zhang et al. \cite{Zhang_2015} deviated from the body of prior work insofar as both focus exclusively on a single port scanning concept. Both studies also demonstrate an original port scanning tool. Yet, similar to De Vivo et al. \cite{de1999review} and Barnett et al. \cite{barnett2008towards}, Kumar and Sudarson provided a technical analysis instead of general principles. To that end, Kumar and Sudarson explored a novel UDP port scanning mechanism. The objective was to increase scanner performance (i.e., time to complete a scan). Zhang et al. \cite{Zhang_2015} introduced a SYN scan modification to bypass firewall filters. A standout against the backdrop of the literature, the authors presented a rigorous experiment and protocol to test hypotheses. Neither study measured accuracy, false positives, false negatives, or efficiency though.

\subsection{Port Scanner Efficacy}
Taxonomies and other modes of knowledge organization are vital to the development of a scientific field. At the same time, the significance of such taxonomies is limited to researchers. Applied experimentation and analysis is required for practical use cases. Fortunately, a few studies \cite{im2016performance, Yuan_2020} recognized the need for future work to demonstrate port scan results in more detail, such as the accuracy.

Im et al. \cite{im2016performance} analyzed the accuracy and performance of network scanning, specifically Nmap. Im et al. defined accuracy as the ability to correctly identify the target's operating system. Then, the authors used an experimental design to capture data while scanning a lab network environment with more than 40 network devices running Windows and GNU/Linux. Overall, they found Nmap had a 27.5\% scan accuracy with 47.8\% scan precision while targeting Windows 7 SP1 with a firewall active. Without a firewall present, scan accuracy and precision jumped to 95\% and 100\% respectively. For GNU/Linux, with the firewall actively filtering traffic, Nmap demonstrated zero percent accuracy and precision. Absent a firewall, Nmap produced 45\% scan accuracy with 100\% precision. 

While these results are important contributions to the the field, even more compelling are the captured frequencies of true positives (TP), false positives (FP), and false negatives (FN). These metrics were collected by Im et al. \cite{im2016performance} on two dimensions: Windows 7 SP1 and GNU/Linux and firewall versus no firewall. In the case of Windows targets, Nmap generated 11 TP, 12 FP, and 17 FN with a firewall active and 38 TP, 0 FP, and 2 FN without the firewall. Against GNU/Linux, Nmap had zero TP and FP with 40 FN while a firewall was active. Without the firewall, Nmap generated 18 TP, zero FP, and 2 FN.

Yuan et al. \cite{Yuan_2020} found Nmap, Zmap, masscan, and unicornscan had defects making each not suitable for practical application. Thus, the authors developed a custom port scanning solution within the constraints of a specific use case. Then, the custom port scanner was run against a production network concurrent to running the other port scanners. Results included performance of each scanner (the same metric as Staniford et al. \cite{Staniford2002Scan} concept of script) with repeated experimental trials in both vertical and horizontal modalities. Overall, the data suggest the custom port scanner developed by Yuan et al. \cite{Yuan_2020} was significantly faster than the other tools. However, the authors did not indicate whether their solution had significantly different rates of false positives, false negatives, or efficiency. 

\section{Method}
Taking into account the features and gaps in the related work, we became interested in what potential quantitative differences exist between popular port scanning tools. According to McGeoch \cite{mcgeoch2012guide}, this is a common approach to applied experimentation in computer science and related fields (e.g., cybersecurity). From this perspective, we framed the research by posing a single research question in this study. The research question is, \textit{to what extent do Nmap, ZMap, and masscan exhibit different accuracies, false positive and negative rates, and system utilization efficiencies when conducting the same type of port scan}?

To generate a potential answer, we opted for a comparison experiment \cite{Tedre2014} design. The experiment incorporated a set port scanning tools (Nmap, Zmap, masscan) as independent variables. The dependent variables were accuracy, false positives, false negatives, and efficiency. Moreover, we established a broad set of controlled variables as (a) scan system, (b) target endpoints, (c) and active services on target endpoints. These are detailed further in the \textit{Experimental Environment} section. Finally, we operationalized the research question and identified variables into a set of hypotheses as follows:

\begin{itemize}[leftmargin=*]
	\item The null hypothesis ($H_0$) is stated as, \textit{The port scanning tools do not exhibit statistically significant differences in accuracy, false positive rates, and false negatives rates}.
	\item Alternatively, $H_1$ is stated as, \textit{The port scanning tools exhibit statistically significant differences in accuracy, false positive rates, and false negatives rates}.
\end{itemize}

We refer to the above as the \textit{general performance} hypotheses. Additionally, we established a pair of \textit{efficiency} hypotheses:

\begin{itemize}[leftmargin=*]
	\item The null hypothesis ($H_0$) is stated as, \textit{The port scanning tools do not exhibit statistically significant differences in CPU or RAM system utilization and scan runtime}.
	\item Alternatively, $H_1$ is stated as, \textit{The port scanning tools exhibit statistically significant differences in CPU or RAM system utilization and scan runtime}.
\end{itemize}

\subsection{Experimental Environment}
The experimental environment consisted of a port scanning virtual machine, a set of 20 endpoint target virtual machines, all interconnected through a 1GBe physical LAN segment. The virtualization host had an AMD Ryzen 9 5900X processor and 128GB of RAM. The host operating system was Ubuntu Desktop 22.04.1 LTS. Each virtual machine, scanning machine and endpoint targets alike, were configured with a single vCPU, 2GB RAM, and a Intel PRO 1000MT virtual network interface. The virtual machines all ran Ubuntu Server 22.04.02 LTS. 

Virtual machine network interfaces were configured in \textit{bridged} mode but with static IP addressing. Further, we configured the physical LAN segment as a class C network with the scan host residing at 192.168.100.10 and the endpoint targets occupying 192.168.100.101 through 192.168.100.120. We connected the virtualization host to an air gapped Cisco 3560 switch running a layer 3 interface in the experimental class C subnet.

\subsection{Hosts and Services}
Port scanning also requires ports to enumerate. Therefore, we installed six services across the 20 scan targets. All targets ran SSH as a control baseline. The remaining five services were distributed across subgroups as detailed in Table 1.

\begin{center}
	\captionof{table}{The list of scan targets and services}
	\begin{tabular}{ccc}
		\hline
		Hosts & Service Group & Service \\ \hline \hline
		.101 - .110 & 80/tcp & HTTP \\ 
		.105 - .114 & 23/tcp & Telnet \\ 
		.108 - .115 & 21/tcp & FTP \\ 
		.111 - .120 & 111/tcp, 2049/tcp & NFS \\ 
		.101 - .120 & 22/tcp & SSH \\ \hline
	\end{tabular}
\end{center}

The intention with this experimental architecture was to facilitate both horizontal (across the set of hosts) and vertical (across a set of services per host). The scan host is a single source for the scans and therefore did not require additional configuration beyond installing the necessary scanner packages.

\subsection{Experimental Procedure}
We developed a procedure to follow during the experiment. While not a direct control, having a step-wise approach ensured consistency throughout the three scanner trials. After powering up the scan host and scan targets, we captured a baseline measure for CPU and RAM baseline on the scan host. Next, we executed an Nmap SYN scan against the experimental subnet. Both the scan results as well as the CPU and RAM utilization were saved to file on the scan host. Then, we reset all systems to clear buffers and caches. The steps from running a port scan to resetting systems were repeated for Zmap and masscan. Finally, we pulled the scan files and utilization profiles from the scan host through USB.  

\subsection{Data Analysis}
We analyzed the collected scan files and utilization profiles using statistical methods to determine if there is a significant difference between the tools in terms of accuracy, false positives, false negatives, and (separately) efficiency. First, we calculated the statistical means for the identified variables under each scan. Thereafter, we conducted an ANOVA test to determine if any difference existed between outcomes followed by an Ad Hoc test if so. 

\section{Results}
We present the results of the data analysis in two sections: \textit{general performance} and \textit{efficiency}. Each section contains a descriptive breakdown of the data as well as the inferential statistics when appropriate.

\subsection{General Performance}
General performance captured how well a given port scanner identifies hosts and services. Identifying all hosts ($N = 20$) and all services ($N=120$) would be one extreme. The opposite extreme would be identifying zero hosts and thus zero services. There were a variety of two dimensional vectors possible in between these extremes. 

\begin{center}
	\captionof{table}{General performance of three port scanners}
	\begin{tabular}{llcccc}
		\hline
		 & N &  Accuracy & FP & FN  \\ \hline \hline
		Nmap & 20 (120) & 100\% & 0 & 0 \\
		Zmap & 20  (120) & 100\% & 0 & 0  \\
		masscan & 20  (120) & 100\% & 0 & 0  \\ \hline
	\end{tabular}
\end{center}

The results show all three port scanners identified the full battery hosts and services correctly (Table 2). Furthermore, no port scanner detected a host or service not present in the experimental configuration. An ANOVA would be superfluous given the extreme condition of the data as the lack of difference between scanners is \textit{prima facie} true. Accordingly, we accept the null hypothesis under these conditions.

\subsection{Efficiency}
We took a snapshot of the scan host CPU and RAM utilization prior to each scanning activity. The system demonstrated  0\% time spent on user processes and 1241652 K in free memory or approximately 1230.7 megabytes in each snapshot. Recall that the virtual machine had a single vCPU allocated along with 2GB of RAM (i.e., 2048 megabytes). Such a baseline suggested the system was appropriately quiescent and had a system overhead of 8173 megabytes. 

We captured efficiency data during the three port scanner trials by writing \Verb|vmstat| output to file in three second intervals. The mean ($M$) was calculated for the user processes CPU elements as well as for the free RAM elements (Table 3). Scan runtimes, in seconds, were calculated through difference between start and end timestamps for each port scan trial.

\begin{center}
	\captionof{table}{Port scanner CPU and RAM system utilization and scan runtime}
	\begin{tabular}{rccc}
		\hline
		& CPU & RAM & Runtime \\ \hline \hline
		Nmap & 1 & 1230656	& 26.68 \\
		Zmap & 1.11	& 988040 & 644 \\
		masscan & 0	& 978492 & 2 \\ \hline
	\end{tabular}
\end{center}

Unlike the observable identical general performance across port scanners, the efficiency measures demonstrated variability.  Thus, we carried out an ANOVA as planned. The statistic revealed there was a difference in efficiency between the scanners ($F$ (between groups $df=2$, within groups $df=6$), $F=153.9721$, $p=0.000006981$). What is more, a Tukey’s HSD test confirmed the means of the variables were significantly different with a large effect size ($7.16$).  We reject the null hypothesis under these conditions.

\section{Conclusion}
Port scanning has a critical role in cybersecurity network defense. The technology allows for enumeration of computing systems on networks and, additionally, active services on those endpoints. Existing literature provides a robust taxonomy for port scanning tools, scan types, and scan techniques. Yet, no work has been done to quantify the efficacy for these popular tools in a uniform, rigorous manner. This study attempted to fill in this gap. To that end, we experimentally compared efficacy of three popular port scanning tools - Nmap, masscan, and Zmap. Specifically, we measured accuracy, false positives, false negatives, and efficiency during single source SYN type port scans using a combined horizontal and vertical technique. The experiment consisted of a virtual scan host and 20 virtual scan targets. The scan targets exposed a mixture of six services- FTP, SSH, Telnet, HTTP, and two ports for NFS. We collected data of each scan trial and from the scan host during each of the scans. The results demonstrated no difference between tools in terms of accurately identifying targets. However, in the case of port scanner efficiency, the results revealed a statistically significant difference.

The results have implications for researchers and practitioners alike. On one hand, researchers and practitioners may take the lack of difference in accuracy between port scanning tools as a signal that algorithms leveraging TCPIP protocol standards are stable. Furthermore, existing implementations such as Nmap, Zmap, and masscan can server as baselines for new tool development. Future work might turn to  unexplored algorithmic territories to expand scanner stability while decreasing technique detection. On the other hand, the significant differences between port scanner efficiency paints a clear path forward for research-based and practical innovation. An additional idea therein might be to expand the concept of \textit{efficiency} to include green computing key performance indicators. 

Of course, no recommendation is without caveat. For instance, while Nmap can use a variety of scan techniques, the other tools used in this experiment use SYN scanning exclusively. Accordingly, the comparative analysis compared scans using that technique only. Doing so limits the generalizability of the results because other types of scans cannot be evaluated (i.e., ACK, FIN, XMAS, and so forth).  Further considering generalizable outcomes, we assume Nmap, Zmap, and masscan implement SYN scanning in similar enough fashion for the comparative analysis to hold. Because SYN scanning follows a standard scanner to client interaction pattern, it is reasonable to conclude the SYN scan implementation is more-or-less identical. At last, future work may be beneficial if deep technical analysis at the protocol level confirmed such interactions.

On that note, we did observe one anomaly during the experiment. We found Zmap and masscan did not function as expected on a closed host-only VMWare network. Future work should explore this more deeply. For now, we observed scan packets leaving the scan host as well as being received on the scan targets. This was confirmed through \Verb|tcpdump| packet capture. Our best speculation is the scan targets responded outside of the port scanners' timeout windows. Notably, Nmap worked identically on the bridged LAN and the closed host-only configuration. Future work may be necessary to explore the technical underpinnings of this phenomenon.

\bibliographystyle{unsrt}
\bibliography{references}

\end{multicols*}
\end{document}